# Fabrication, properties, and applications of flexible magnetic films [*]

Liu Yiwei (刘宜伟), Zhan Qingfeng (詹清峰)[†], and Li Run-Wei (李润伟)[†]

Key Laboratory of Magnetic Materials and Devices, Ningbo Institute of Material Technology and Engineering (NIMTE), Chinese Academy of Sciences (CAS), Ningbo 315201, People's Republic of China.

**Abstract:**

Flexible magnetic devices, i.e., magnetic devices fabricated on flexible substrates, are very attractive in application of detecting magnetic field in arbitrary surface, non-contact actuators, and microwave devices due to the stretchable, biocompatible, light-weight, portable, and low cost properties. Flexible magnetic films are essential for the realization of various functionalities of flexible magnetic devices. To give a comprehensive understanding for flexible magnetic films and related devices, we have reviewed recent advances in the studies of flexible magnetic films including fabrication methods, magnetic and transport properties of flexible magnetic films, and their applications in magnetic sensors, actuators, and microwave devices. Three typical methods were introduced to prepare the flexible magnetic films. Stretching or bending the flexible magnetic films offers a good way to apply mechanical strain on magnetic films, so that magnetic anisotropy, exchanged bias, coercivity, and magnetoresistance can be effectively manipulated. Finally, a series of examples were shown to demonstrate the great potential of flexible magnetic films for future applications.

[*]Project supported by the National Natural Foundation of China (11274321, 11174302, 11374312, 11304326), the State Key Project of Fundamental Research of China (2012CB933004, 2009CB930803), the Chinese Academy of Sciences (CAS), and the Ningbo Science and Technology Innovation Team (2011B82004, 2009B21005), and the Zhejiang and Ningbo Natural Science Foundations (2013A610083).

[†]Corresponding author. E-mail: zhanqf@nimte.ac.cn and runweili@nimte.ac.cn

# 1. Introduction

Silicon wafers have been widely used in manufacturing electronic devices [1]. However, many natural things, such as human bodies, organisms, clothes, etc., are elastic, soft, and curved [2]. Therefore, the electronic devices based on rigid silicon wafers are not suitable for these new coming applications. Flexible electronics is a technology of assembling electronic circuits and devices on flexible substrates, which are significantly of lower cost, lighter, and more compact, as compared to the conventional electronic devices [3]. By thinning single crystal silicon wafer to 100 μm, the first flexible solar cell was made in 1960 [4,5], which started the development of flexible electronics. In 1997, polycrystalline silicon thin film transistor (TFT) made on plastic substrates was reported and received lots of attentions because of its potential applications in flexible display [6,7]. Since then, flexible electronics has been developed rapidly, and commercialized products have appeared in our daily life. Now, flexible electronic devices have been widely used in display, radio-frequency identification (RFID), solar cell, lighting, and sensors, among which flexible displays occupied more than 80% of the market of flexible electronics [8,9,10]. With the development of various flexible devices, display, logic, sensor, and memory are expected to be integrated in a multi-functional system and be fabricated on a flexible substrate [11].

It is well known that magnetic materials are important for fabrication of electronic devices [12]. For example, soft magnetic materials are usually applied in inductors, transformers, microwave devices, and screening of magnetic field [13,14]. Hard magnetic materials are widely used in loudspeakers, generators, memories, and sensors [15]. Magnetostrictive materials are used in actuators [16]. Permalloy can be used in anisotropic magnetoresistance (AMR) sensors [17]. Giant magnetoresistance (GMR) or tunneling magnetoresistance (TMR) multilayered structures can be used in high-speed read/write heads in disk memory devices owning to their large magnetoresistance [18,19]. Recently, GMR or TMR sensors fabricated on flexible substrates, so called flexible magnetoelectronics, have attracted a lot of interests due to their potential applications in detecting magnetic field in living organisms [20,21]. Because of the extensive applications of the magnetic materials, it is inevitable to integrate flexible magnetic materials in flexible electronics.

In this review, we first introduced the techniques of fabricating flexible magnetic films and devices. Then, we focused on the properties of flexible magnetic films and devices. Finally, the applications of flexible magnetic films and devices were discussed.

## 2. Fabrication of flexible magnetic films

### 2.1 Magnetic films deposited on flexible substrates

Fabricating magnetic films directly on flexible substrates is a straightforward way to get flexible magnetic films. The most used flexible substrates are organic polymers including polyethylene terephthalate (PET), polyethylene naphthalate (PEN), polyethersulphone (PES), polyimide (PI), and polydimethylsiloxane (PDMS) [22]. These organic polymers are highly flexible, inexpensive, and compatible with the roll-to-roll processing [23]. Most of the polymers cannot suffer a high-temperature treatment [24], but they are still suitable for fabrication of most of magnetic films and devices because of the near room temperature deposition and low temperature (usually lower than 400 $^o$C) post-annealing.

For fabricating magnetic films on flexible substrates, a suitable buffer layer is often required to reduce the roughness of flexible substrates and ensure the continuity and functionality of magnetic films and multilayered structures, such as GMR and spin-valve devices [25,26,27]. For example, the root-mean-square (RMS) roughness of a PET substrate is about 2.16 nm, which is much larger than that of the thermally oxidized Si substrate. The 150-nm-thick $Fe_{81}Ga_{19}$ film directly grown on flexible PET substrates exhibits a RMS roughness of 3.34 nm[28]. A 30-nm-thick Ta buffer layer can reduce the roughness of $Fe_{81}Ga_{19}$/Ta/PET films to 2.04 nm [29]. By growing a Ta buffer layer, flexible exchange biased Ta(5 nm)/$Fe_{81}Ga_{19}$(10 nm)/$Ir_{20}Mn_{80}$(20 nm)/Ta(30 nm)/PET heterostructures were successfully fabricated, which could be used to stabilize the magnetization of magnetic layers in flexible spin-valve devices [30]. Chen *et al.*, have fabricated flexible Co/Cu GMR multilayers on polyester substrates by dc magnetron sputtering [31]. The sample structure is schematically shown in Fig. 1(a). Figure 1(b) shows the photographic image of circularly bended Co/Cu multilayer deposited on polyester substrates. Before deposition the Co/Cu multilayer, AR-P 3510 positive photoresist (Allresist, Germany) buffer layer with the thickness of 2 μm has been spin-coated on flexible substrates to reduce the surface roughness.

Besides, Oh *et al.*, fabricated flexible spin-valve structures of Ta (3 nm)/NiFe (10 nm)/Cu (1.2 nm)/NiFe (3 nm)/IrMn (10 nm)/Ta (3 nm) on PEN substrates using AZ 5214E photoresist as the buffer layers [32]. Melzer *et al.*, provided another way to fabricate flexible spin-valve structures, as shown in Fig. 2 [33]. First, PDMS was spin-coated onto silicon wafers with surface roughness < 0.5 nm. Then, the structure of Ta (2 nm)/IrMn (5 nm)/ [Permalloy (Py) (4 nm)/CoFe (1 nm)]/Cu (1.8 nm)/[CoFe (1 nm)/Py (4 nm)] with 5-nm-thick Ta buffer layer was fabricated by magnetron sputtering. After a lithographic lift-off process, by means of the antistick layer, the PDMS film was peeled from the rigid silicon wafer, forming a flexible magnetic multilayered structure.

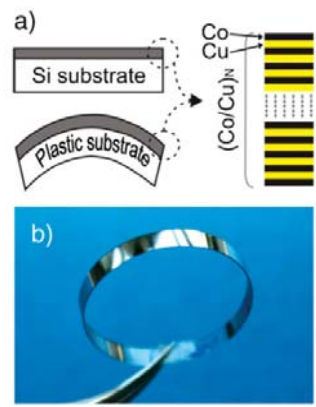

**Fig. 1.** (a) Schematic illustration of $(Co/Cu)_N$ films deposited on Si and flexible substrates. (b) A photographic image of circularly bended $(Co/Cu)_{20}$ film deposited on polyester substrate.[31]

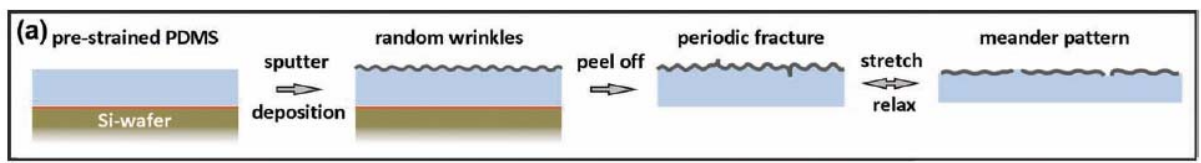

**Fig. 2.** Fabrication process of stretchable spin-valve structure. [33]

with the development of magnetoelectric materials, multiferroic composites consisting of magnetostrictive materials and organic ferroelectric materials, such as polyvinylidene fluoride (PVDF) and polyvinylidene-fluoride–trifluoroethylene (PVDF-TrFE), have received much attention [34]. Sm–Fe/PVDF heterostructural films have been prepared by depositing Sm–Fe nanoclusters onto flexible PVDF membranes using cluster beam deposition, which exhibits large magnetoelectric voltage output of 210 μV at an external magnetic bias of 2.3 kOe[34].

Besides organic polymers, people also seek for other types of flexible substrates to prepare magnetic films. Liang *et al.*, have successfully produced flexible graphene/$Fe_3O_4$ hybrid papers by using graphene as flexible substrates, as shown in Fig. 3 [35]. During fabrication, a two-step process was used: (I) mixing of graphene aqueous solution with water-soluble $Fe_3O_4$ nanoparticles and (II) chemical reduction of the suspension of water-soluble $Fe_3O_4$ nanoparticles and graphene sheets with hydrazine. In a broad sense, cantilevers with several micrometer in thickness, which have been applied as sensors in micro-electro-mechanical systems (MEMS) [36], can be employed as flexible substrates. The cantilevers are usually made of materials including Si, polyimide, $Si_3N_4$, etc [37]. Onuta *et al.*, have fabricated a flexible multiferroic composite as an energy harvester consisting of a magnetostrictive $Fe_{0.7}Ga_{0.3}$ thin film and a $Pb(Zr_{0.52}Ti_{0.48})O_3$ piezoelectric thin film on a 3.8-μm-thick Si cantilever, as shown in Fig. 4 [38]. Magnetic polymers fabricated by dispersing magnetic components in polymer matrix, can be served as flexible functional cantilevers, which can respond to external magnetic fields and mechanical vibrations. If the magnetic components are magnetostrictive materials and the polymers are ferroelectric materials, three main types of polymer-based multiferroic materials, nano-composites, laminated composites, and polymer as a binder composite can be achieved, as shown in Figs. 5 (a), 5(b), and 5(c), respectively[39]. The investigations on polymer-based multiferroic materials are challenging and innovative, which bridge the gap between fundamental research and applications in the near future.

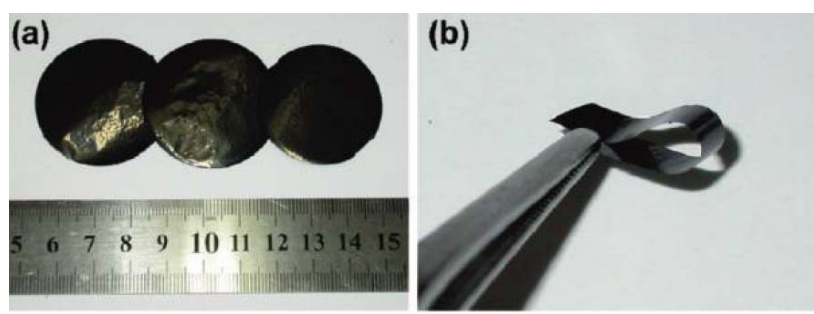

**Fig. 3.** The flexible graphene/$Fe_3O_4$ hybrid papers using graphene as the flexible substrates. [35]

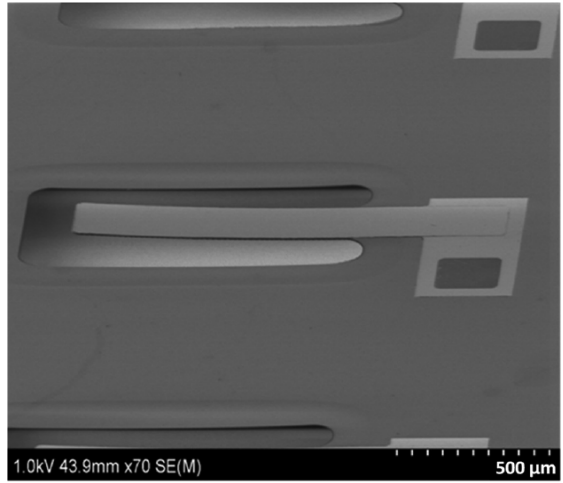

**Fig. 4.** A scanning electron micrograph (SEM) of a flexible multiferroic energy harvester.[38]

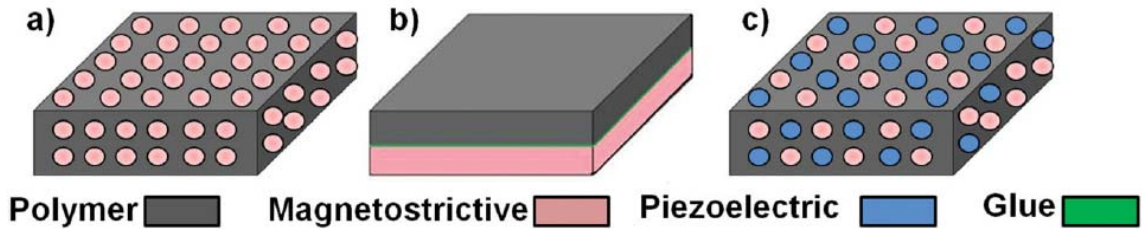

**Fig. 5.** Three main types of Polymer-based multiferroic materials. [39]

## 2.2 Transfer and bonding approach

In transfer and bonding approach, magnetic films or structures are fabricated on conventional rigid substrates like Si wafer, glass, MgO, etc., by standard fabrication methods. Then, the magnetic films or structures can be transferred by removing the substrates through laser annealing, chemical solution, or directly peeling off the films [40,41,42]. Finally, the transferred films or structures can be bonded to flexible substrates by glue or physical/chemical adsorption [40,41,42]. Generally, conventional rigid substrates are much more flat than flexible substrates. Therefore, the transfer and bonding approach can provide flexible magnetic films and structures with high quality and high performance. However, it is difficult to prepare a large area flexible film by means of this method due to the random damage during the transfer procedure. After the discovery of graphene, a variety of transfer and bonding approaches have been developed to transfer graphene films onto flexible substrates [43]. However, these methods cannot be directly employed to prepare flexible magnetic films, because some chemicals used in the transfer procedure may damage the magnetic films [44]. Donolato *et al.*, have developed another innovative, simple, and versatile pathway to transfer magnetic films and structures onto flexible polymer substrates, as shown in Fig. 6 [45]. At first, a trilayered structure of Ti/Au/$SiO_2$, which acts as the donor substrate, was grown onto a conventional Si substrate using a sputter deposition system. Then, magnetic nanostructures made of permalloy ($Ni_{80}Fe_{20}$) were fabricated on the trilayer by means of an electron beam lithography process. After that, PDMS was spin coated on the magnetic nanostructure. Finally, the transfer was accomplished by simple immersion of the chip in water and a gentle mechanical lifting of the polymer membrane off the substrate. In this approach, the selection of Au and $SiO_2$ layers is the key aspect. Due to the hydrophobic character of Au and hydrophilic behavior of $SiO_2$, the bond between them is rather weak, so the Au and $SiO_2$ layers can be easily separated via the water-assisted lift-off process.

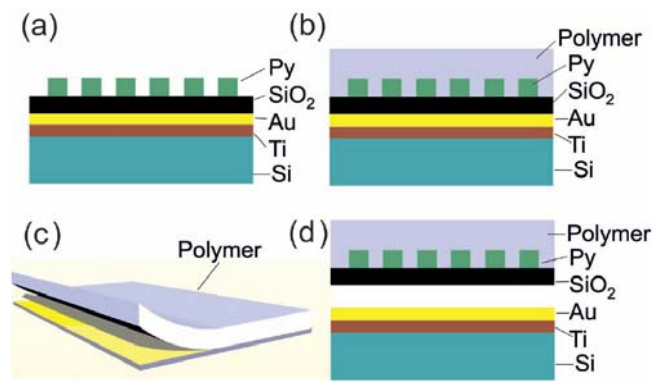

**Fig. 6.** Schematic show of the transfer and bonding approach to get flexible magnetic films. [45]

**2.3 Release of sacrificial layers**

The general processes of releasing sacrificial layers for preparing flexible magnetic films are shown in Fig. 7. Magnetic films are first deposited on bulk substrates or bulk substrates with sacrificial layers. For the simple film/substrate structures, the substrates themselves can be treated as sacrificial layers. The substrates or the sacrificial layers can be removed by aqueous solution of chemicals, chemical etching, or dry etching to achieve freestanding magnetic films [46,47]. This method has the similar advantage as the above-mentioned transfer and bonding approach and can provide flexible magnetic films with high quality.

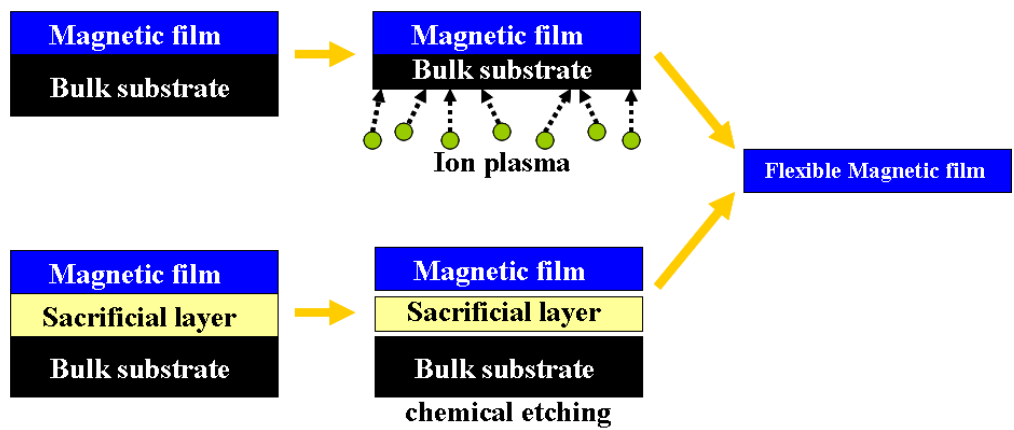

**Fig. 7.** The general processes of releasing sacrificial layers for preparing flexible magnetic films.

The most used sacrificial layers are NaCl and photoresist. Heczko and Thomas have epitaxially grown Ni-Mn-Ga films on water-soluble (001)-oriented NaCl single crystals and obtained high quality free-standing Ni-Mn-Ga films by dissolving the NaCl substrates[48]. On the other hand, Tillier *et al.*, have used photoresist as sacrificial layers to fabricate flexible Ni-Mn-Ga films [49]. Ni-Mn-Ga films were first deposited on photoresist (Shipley S1818) layers, which have been spin-coated on polycrystalline $Al_2O_3$ substrates. After deposition, the samples were placed in an acetone bath to remove the photoresist sacrificial layer and obtain the freestanding magnetic films. Other materials, such as Au, MgO, and Cr, can also be used as sacrificial layers due to their chemical soluble properties [50,51,52]. For example, Bechtold *et al.*, have prepared 1.2-μm-thick $Fe_{70}Pd_{30}$ films on Au(50 nm)/Cr(8 nm)/MgO substrates [50]. After deposition, the $Fe_{70}Pd_{30}$ films were released from the substrate by wet chemical etching of the sacrificial Au layer in an aqueous solution of potassium iodide and iodine. Although removing sacrificial layers is a good way to prepare high quality flexible magnetic films, this method still has some disadvantages. For example, the aqueous solution of chemicals may damage the magnetic films. Alternatively, flexible membranes of inert materials, such as Pt and Au, can be prepared by removing sacrificial layers. Then magnetic films can be deposited on the thin metallic membranes. We have etched platinized Si substrates (Pt(200 nm)/Ti(50 nm)/$SiO_2$(500 nm)/Si) in 10 wt% HF solutions for 4 h[53]. Because Ti and $SiO_2$ layers reacted with HF, producing soluble $SiF_4$ and $TiF_3$, respectively, the 200-nm-thick Pt layers were released from Si substrates. The flexible Pt foils can be used as flexible substrate for preparing ferromagnetic or ferroelectric films at an elevated temperature. The detailed processes are schematically shown in Fig. 8.

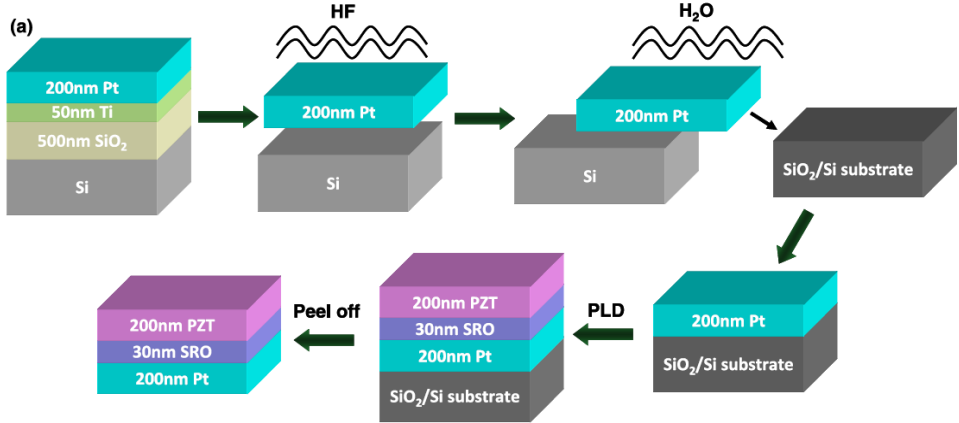

**Fig. 8.** Method to obtain flexible functional thin-films using Pt foils. [53]

## 3. Properties of flexible magnetic films

### 3.1 Effect of buffer layer

Prior to fabricating flexible magnetic films and devices, an appropriate buffer layer needs to be introduced to decrease the roughness of flexible substrates, improve the crystal orientation of magnetic films, and release the residual stress. Therefore, the buffer layers are extremely important in determining the properties of flexible magnetic films, such as magnetic anisotropy, coercivity, magnetoresistance, etc. We have investigated the effect of a Ta buffer layer on the magnetic properties of magnetostrictive $Fe_{81}Ga_{19}$ films grown on flexible PET substrates [29]. As shown in Fig. 9, with increasing the thickness of Ta buffer layer, both the uniaxial magnetic anisotropy and coercivity of $Fe_{81}Ga_{19}$/Ta/PET films are decreased. Obviously, the Ta buffer layer could effectively release the residual stress in PET substrates, and therefore reduce the strength of the uniaxial anisotropy of $Fe_{81}Ga_{19}$ layers. The decrease of coercivity of $Fe_{81}Ga_{19}$ films may result from both the decrease of uniaxial anisotropy and the flatness of the films. Chen *et al.*, have shown that the GMR effect of Co/Cu multilayers (MLs) on a flexible organic substrate can be enhanced up to 200% by introducing a photoresist (PR) buffer layer to flatten the plastic substrates [31]. They have compared three Co/Cu multilayers grown on Si substrate, polyester substrate (P), and polyester substrate with 2 μm photoresist (PR) buffer layer (PR+P). As shown in Fig. 10(d), the RMS roughness, Rq, of the P substrate is one order of magnitude larger than that of Si substrate, and Rq of the PR+P substrate has a similar value obtained for the Si substrate. As shown in Figs. 10(a), 10(b), and 10(c), GMR values significantly increase after introducing a PR buffer layer and rise to even higher values than those achieved on Si substrates due to an increased antiferromagnetic coupling fraction of the flexible PR buffered Co/Cu multilayers.

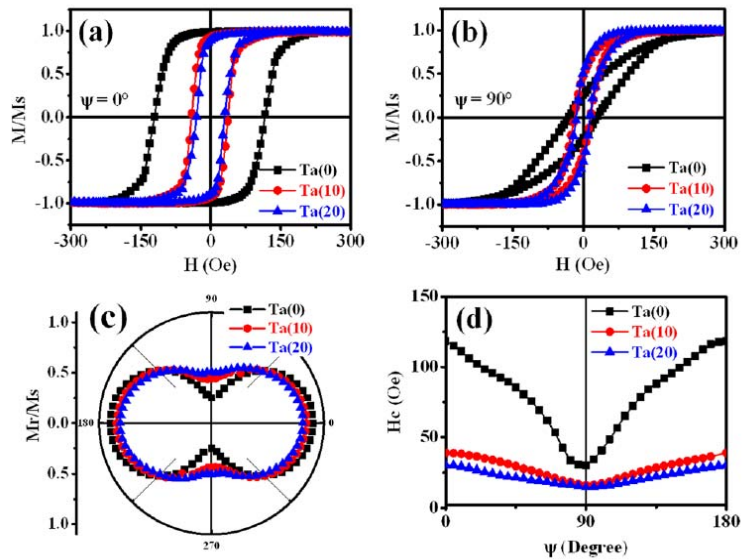

**Fig. 9.** Hysteresis loops for flexible $Fe_{81}Ga_{19}$(50 nm)/Ta/PET films with a magnetic field applied along (a) the easy ($\psi=0°$) and (b) hard axes ($\psi=90°$), and the corresponding angular dependence of (c) squareness and (d) coercive field. Ta(0), Ta(10), and Ta(20) indicate Ta buffer layer with thickness of 0, 10, and 20 nm, respectively. [29]

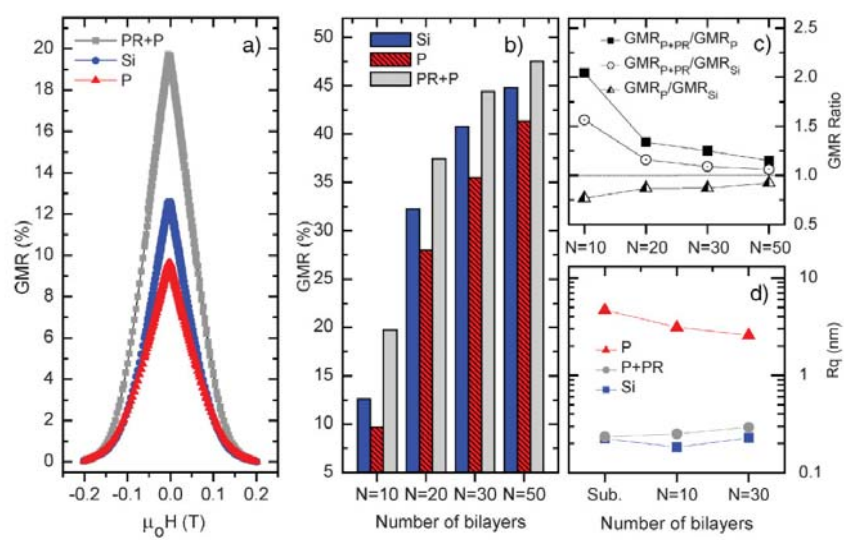

**Fig. 10.** (a) GMR curves of $(Co/Cu)_{10}$ MLs deposited on Si, P, and PR+R substrates. (b) GMR comparison of $(Co/Cu)_N$ MLs on various substrates with different numbers of bilayers. (c) GMR ratio of $(Co/Cu)_N$ MLs deposited on Si, P, and PR+P substrates. (d) RMS roughness, Rq, of Si, P, PR+P bare substrate and Co/Cu films on corresponding substrates. [31]

**3.2 Strain dependence of magnetic properties**

Control of magnetic properties of flexible magnetic films via mechanical strains is an interesting topic from the viewpoint of both fundamental researches and potential applications [54]. Magnetic anisotropy is a key characteristic in determining the direction of magnetization and affecting the performance of spintronic devices [55]. For flexible spintronic devices applied in curved surfaces or used to evaluate the mechanical strain, their magnetic anisotropy under various mechanical strains need to be known and well controlled. FeGa magnetostrictive alloys exhibiting moderate magnetostriction (~ 350 ppm for Ga content of 19%) under very a low magnetic field (~ 100 Oe) but good mechanical properties is a potential material applied in strain controllable spintronic devices. We have fabricated magnetostrictive $Fe_{81}Ga_{19}$ films on flexible PET substrates [28]. Due to the residual stress of the flexible substrates, a uniaxial magnetic anisotropy is observed for the as-grown $Fe_{81}Ga_{19}$ films. By inward or outward bending the PET substrates, the compressive and tensile strains can be applied on $Fe_{81}Ga_{19}$ films. The hysteresis loops for $Fe_{81}Ga_{19}$/PET films under tensile and compressive strains are measured by bending the substrates along the easy or hard axes of $Fe_{81}Ga_{19}$ films, as shown in Fig. 11 [28]. For the magnetic field oriented along the easy axis, a tensile strain along the hard axis gives rise to a drastic decrease in Mr/Ms ratio, as shown in Fig. 11(a). In contrast, under a compressive strain, the Mr/Ms ratio is increased, as shown in Fig. 11(b). For the magnetic field oriented along the hard axis, the Mr/Ms ratio is decreased and increased under a tensile and compressive strain applied along the easy axis, respectively, as shown in Figs. 11(c) and 11(d). The results provide an alternative way to mechanically tune magnetic properties, which is particularly important for developing flexible magnetic devices.

In addition, we have also studied the effect of mechanical strain on magnetic properties of flexible exchange biased $Fe_{81}Ga_{19}$/IrMn heterostructures grown on PET substrates [56]. Figure 12 shows the typical results for the in-plane strain dependence of normalized magnetic hysteresis loops of $Fe_{81}Ga_{19}$(10 nm)/IrMn(20 nm) bilayers with strain applied perpendicular or parallel to the pinning direction (PD). The exchange bias field achieves a maximum value of 69 Oe for magnetic field applied along the induced PD and vanishes for magnetic field perpendicular to the PD, as shown in Figs. 12(a) and 12(b). The loop squareness is decreased when a tensile strain is applied perpendicular to the PD with magnetic field parallel to the PD or a tensile strain is applied parallel to the PD with magnetic field perpendicular to the PD. Different from the previously reported works on rigid exchange biased systems, a drastic decrease in exchange bias field was observed under a compressive strain with magnetic field parallel to the PD, but only a slightly decrease was shown under a tensile strain. Based on a modified Stoner-Wohlfarth model calculation, we suggested that the distributions of both ferromagnetic and antiferromagnetic anisotropies be the key to induce the mechanically tunable exchange bias.

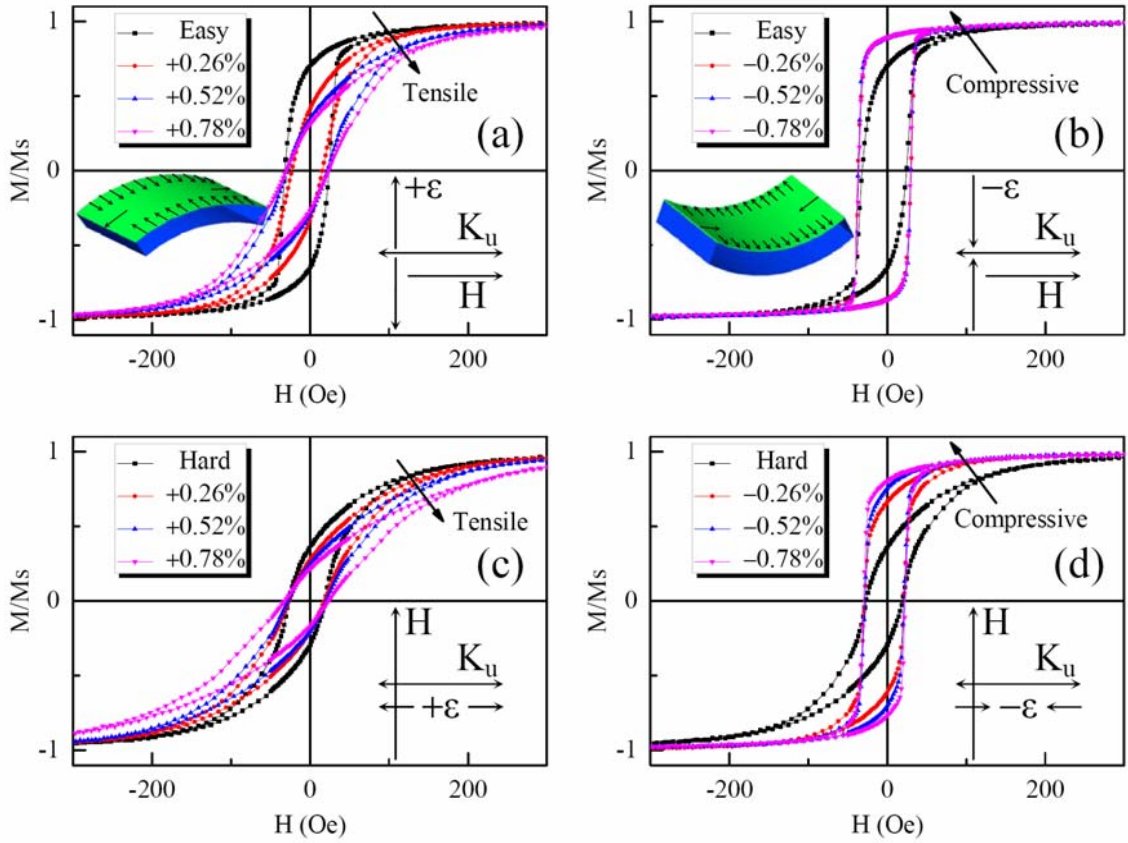

**Fig. 11.** Hysteresis loops for $Fe_{81}Ga_{19}$/PET obtained under various external strains using different measuring configurations, (a) Magnetic field $H$ parallel to the uniaxial anisotropy $K_u$ and a tensile strain $\varepsilon$ (outward bending of PET substrates) applied perpendicular to $K_u$, (b) $H$ parallel to $K_u$ and a compressive strain $-\varepsilon$ (inward bending) perpendicular to $K_u$, (c) $H$ perpendicular to $K_u$ and $\varepsilon$ parallel to $K_u$, and (d) $H$ perpendicular to $K_u$ and $\varepsilon$ parallel to $K_u$. [28]

In order to understand the effect of strain on magnetic properties from the view of microcosmic, it is necessary to image the magnetic domain structures under different strains. Chen *et al.*, have prepared 100 nm Co films on polyester substrates. The magnetic domain structures with and without plastic strains were obtained by Kerr microscopy, as shown in Fig.

13 [57]. It is found that the size of magnetic domains become much larger and ordered when the magnetic field is aligned along the easy axis. A tensile strain applied along easy axis can further increase the size of the domains. When the magnetic field is applied along the hard axis, the density of magnetic domains is increased after applying a tensile strain along easy axis.

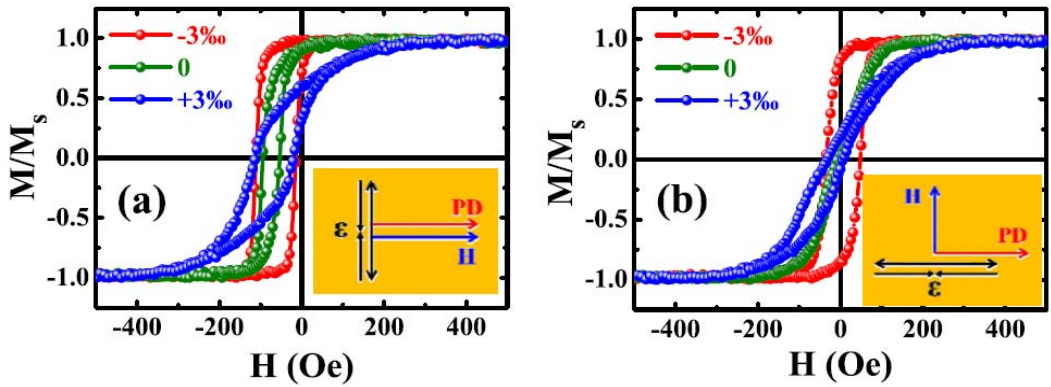

**Fig. 12.** The strain dependence of magnetic hysteresis loops for $Fe_{81}Ga_{19}$(10 nm)/IrMn(20 nm) bilayers with magnetic field (a) parallel and (b) perpendicular to the PD. The compressive and tensile strains are applied perpendicular or parallel to the PD, as shown in the insets of (a) and (b), respectively. [56]

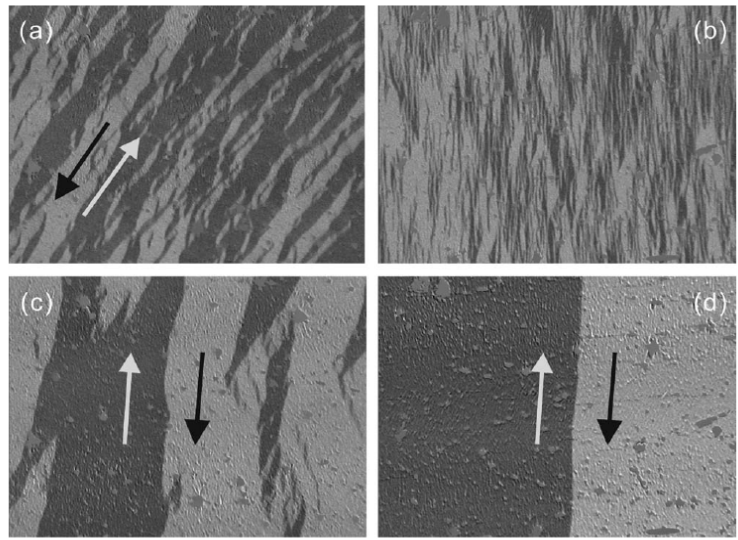

**Fig. 13.** Effect of strain on the magnetic domain structures of 100 nm Co films on polyester substrates.
(a) Strain $\varepsilon$=0 and $H$ along hard axis, (b) Tensile strain $\varepsilon$=0.75% along easy axis and $H$ along hard axis, (c) Strain $\varepsilon$=0 and $H$ along easy axis, (d) Tensile strain $\varepsilon$=0.75% along easy axis and $H$ along easy axis. [57]

### 3.3 Strain dependence of magnetoresistance

Magnetoresistance (MR), where the resistance of the material changes with applied magnetic field, has been extensively used as magnetic field sensors, read-write heads, and magnetic random access memory [19]. As early as 1992, Parkin *et al.*, have prepared flexible GMR multilayers on Kapton substrates, which display 38% room-temperature GMR, almost as large as that found in similar structures prepared on silicon wafers. Such flexible structures suggest potential technological applications in light-weight read heads [58]. In 1996, Parkin also demonstrated flexible exchange-biased magnetic sandwiches with lower saturation field and 3% room-temperature MR suggesting the possibility of manufacturing flexible MR read head [30]. In 2010, Barraud, *et al.*, have successfully prepared flexible Co/Al$_2$O$_3$/Co TMR structure on polyester based organic substrates with 12.5% room-temperature TMR ratio[59]. Since the flexible substrates are easy to distort, due to the magnetostriction effect, the magnetic and transport properties of flexible magnetic films and devices strongly depend on the strain status of samples. How to make the magnetic films and devices insensitive to the strain is a challenging task. Melzer *et al.*, demonstrated an easy approach to fabricate highly elastic spin-valve sensors insensitive to the strain on flexible PDMS [33]. By means of a predetermined periodic fracturing mechanism and random wrinkling, meander-like self-patterning devices can be achieved, as seen in Fig. 2. This meander-like structure makes the device insensitive to the strain. The influence of strain on GMR and sample resistance is shown in Fig. 14 [33]. It is clearly seen that both GMR magnitude and sample resistance maintain their values very stably under different strains, which are suitable as stretchable sensors to detect magnetic field.

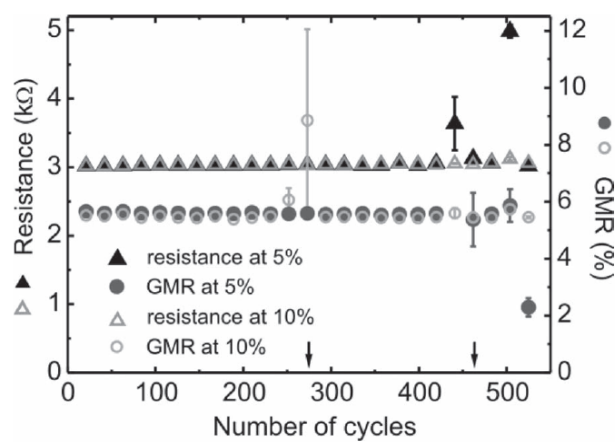

**Fig. 14.** GMR magnitude (circles) and sample resistance (triangles) during a cyclic loading experiment between 5% (filled symbols) and 10% (open symbols) strain. [33]

On the other hand, if MR devices are designed to evaluate strain or controlled by strain, the MR of devices is required to be sensitive to the strain. Such strain tunable MR devices can be used in novel straintronic devices, which consume extremely low power. Generally, a GMR film consists of periods of two ferromagnetic layers separated by a conducting layer, and the MR depends on the relative magnetization directions of the adjacent ferromagnetic layers and the interlayer exchange coupling [60]. An in-plane tensile strain can reduce the thickness of the spacer layers in a GMR structure, resulting in the change of interlayer exchange coupling and the MR, as Chen *et al*., have reported [31]. Another approach for strain control of MR is to tune the magnetic anisotropy of ferromagnetic layers. As mentioned in section 3.2, the magnetic anisotropy of magnetostrictive materials can be manipulated by mechanical strain or stress. Since most ferromagnetic materials exhibit the magnetostriction effect, the mechanical strain control of MR can be realized through the strain-induced magnetic anisotropy [61,62]. Özkaya *et al*., have prepared Co(8 nm)/Cu(4.2 nm)/Ni(8 nm) pseudo-spin-valve (PSV) structures on flexible PI substrates [63]. The low GMR magnitude for the as-prepared sample at zero-field indicates that the magnetization directions of Co and Ni are parallel at the remanent state, as shown in Fig. 15 [63]. Applying a uniaxial strain may lead to opposite rotation of the magnetization directions in both magnetic layers to each other due to different signs of the magnetostriction coefficients of Co and Ni. Upon applying a strain perpendicular to the easy axis of Co and Ni, the zero-field GMR magnitude increases with increasing strain, as shown in Figs. 15 (b), 15(c) and 15(d), which indicates that the angle between the magnetizations of each layer in the remanent state increases. This result suggests that both the magnetic field sensitivity and the magnetic field operating range of GMR devices can be optimized via applying strain.

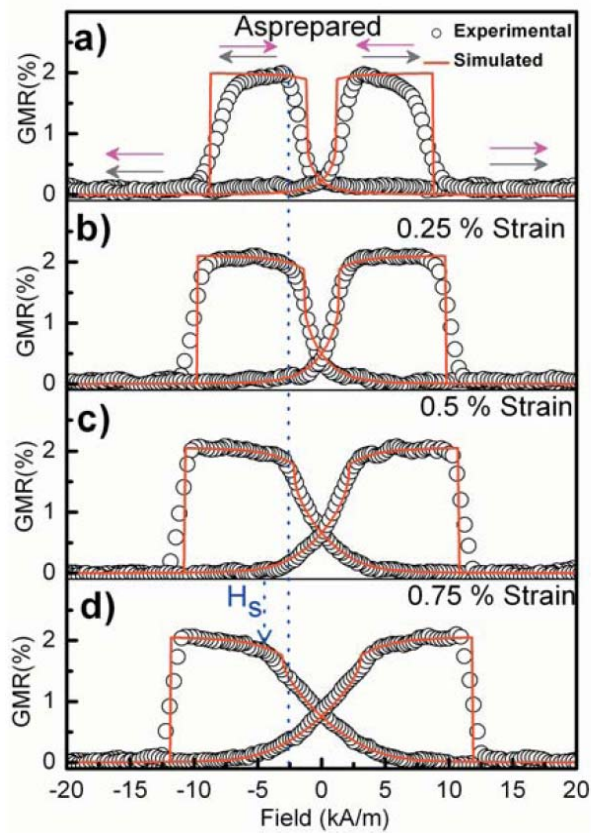

**Fig. 15.** GMR magnitude for Co(8 nm)/Cu(4.2 nm)/Ni(8 nm) GMR structures in
(a) as-prepared state, (b) 0.25%, (c) 0.5% and (d) 0.75% strained states. [63]

## 4. Applications

**4.1 Flexible spintronic devices applied in biomedical techniques**

Magnetic particles can be used to deliver drug or gene, and help to detect proteins, nucleic acids, or to enhance magnetic resonance imaging (MRI) contrast, all of which are very important for the modern biomedical techniques [64,65,66]. In a biomedical system, monitoring and analyzing the signals from magnetic particles are the essential issue, which increases the demand for integration of magnetic field sensing devices into biomedical systems. In this respect, spintronic devices, such as GMR and TMR sensors, provide an efficient solution for detecting magnetic particles due to their high magnetic field sensitivity [67,68]. However, magnetic particles usually flow in the micro-fluidic channels, so that integrating the magnetic sensors with the micro-fluidic channels can significantly improve the sensitivity for detecting magnetic particles. Mönch *et al.*, have successfully fabricated a fully integrative rolled-up GMR sensor simultaneously acting as a fluidic channel for in-flow detection of magnetic particles, as shown in Fig. 16 [69]. This flexible and rolled-up GMR sensor leads to better signal-to-noise ratio and magnetic particles in a fluidic channel can be easily detected and counted. A small disadvantage of this rolled-up GMR sensor is that it requires intensive lithography processing and is therefore expensive and time consuming. Melzer *et al.*, provide another way to prepare the low-cost flexible GMR sensors used in detecting magnetic particles. The preparation process is shown in Fig. 2 [33]. After preparation, the optimized GMR sensors are wrapped around the circumference of a Teflon tube, as shown in Fig. 17(a)

[70]. Figures 17(b) and 17(c) demonstrate the sensor's output, when the magnetic particles are passing through the flexible GMR sensors [70].

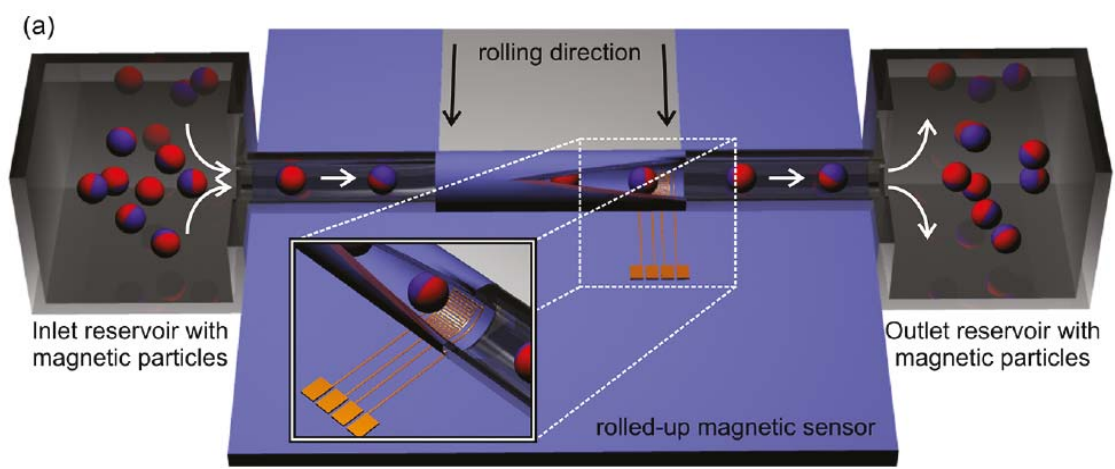

**Fig. 16.** Schematics revealing the main concept of rolled-up magnetic sensor for in-flow detection of magnetic particles. [69]

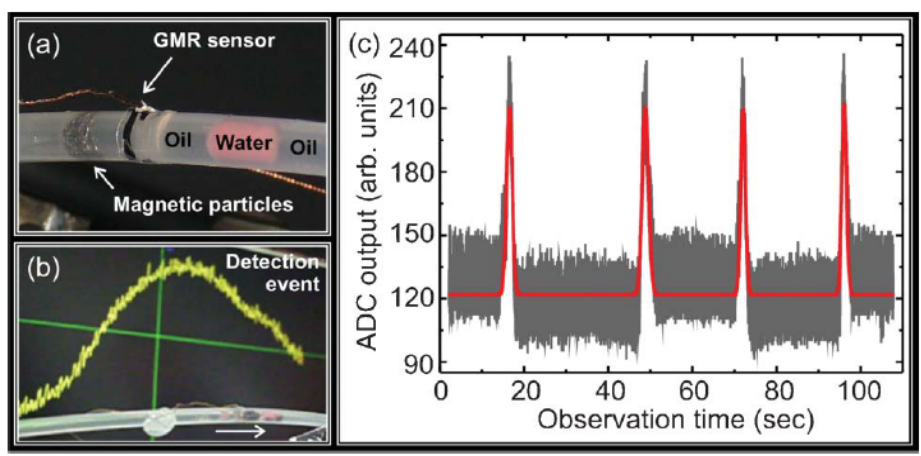

**Fig. 17.** Detection of magnetic particles in a fluidic channel: (a) Elastic GMR sensor wrapped around the circumference of a Teflon tube. The magnetic particles are approaching the GMR sensor. (b) Signal of the elastic GMR sensor on a screen (background) as the magnetic cluster is passing the sensor (foreground). (c) Several consecutive detection events of particles passing the elastic GMR sensor. [70]

The flexible electronic circuits used in the biomedical systems are generally prepared by means of a new-developed printing method, which could revolutionize large area and low-cost electronics manufacturing [71]. However, the fabrication of printable magnetic sensors remains challenging due to lack of magnetic inks containing various components. Karnaushenko *et al.*, have for the first time developed a kind of magnetic ink with GMR flakes which can be easily printed on various substrates, such as paper, polymer and ceramic

[72]. In order to fabricate the magnetic ink, GMR sensors are first deposited on 3 inch silicon wafers with photosensitive polymer AR-P 3510 as the buffer layer. After preparation, the samples are rinsed in acetone to release the deposited GMR sensors from the substrates. The obtained GMR sensors show flake-like or rolled-up structures arising from the intrinsic strain of the deposited GMR films on rigid substrates, as shown in Fig. 18(a) [72]. To assure high electrical conductivity of as-prepared GMR flakes, a multilayer stacked structure is prepared from the originally obtained material by a ball milling. The resulting powder is filtered through a grid that defines the maximum lateral size of a GMR flake to about 150 μm, as shown in Fig.18 (b). A GMR ink is prepared by mixing 500 mg of the GMR powder with 1 ml of a binder solution that is an acrylic rubber based on poly (methyl methacrylate) (PMMA) dissolved in a methyl isobutyl ketone. Finally, using a brush the solution is painted on different surfaces, *i.e.* paper, polymers and ceramic. As shown in Fig.18 (c), the cross-section scanning electron microscopy (SEM) image of a large-area film shows continuous layer stacked structures, which significantly facilitate the electron transport along the in-plane direction to achieve a high GMR effect. This method uses standard sputter deposition, milling and mixing machines for high yield production, demonstrating the suitability of the printable magnetoelectronic devices for large scale industrial production.

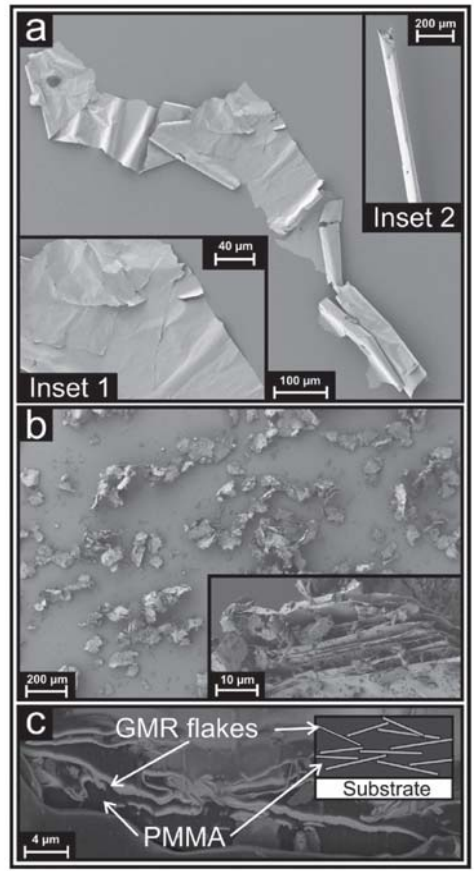

**Fig. 18.** SEM images of GMR powder on various stages of ink preparation: (a) Initial GMR powder directly after delamination from Si substrates that consists of large metallic flakes (inset 1) and a variety of tube like structures (inset 2) self-assembled by releasing of film intrinsic stress. (b) The magnetic film is milled using ceramic beads in order to produce magnetic powder consisted of variously shaped flakes. (c) SEM image of the cross-section through printed sensor shows the internal structure of metallic flakes percolated inside polymer; (inset) schematic drawing demonstrate the principle of flake percolation.[72]

**4.2 Flexible multiferroic structure applied in energy harvesting**

Energy harvesting is a process by which energy can be captured from external sources, such as solar power, thermal energy, vibrational energy, electromagnetic waves, etc, and converted into electrical energy [73]. Energy harvesting technologies can be a substitute for batteries, minimizing the power consumption. Among various energy harvesting technologies, there has been significant interest in the area of vibration energy based on piezoelectric and magnetic harvesters [74,75]. Figure 19 shows the magnetoelectric energy harvesting mechanism of multiferroic composites made by combining magnetostrictive and piezoelectric phases together [39]. First, an external magnetic field leads to deformation the magnetostrictive phase through the magnetostriction effect, and the deformation could be transmitted to the piezoelectric phase across the interface between the two phases, which would generate the electric charges due to the converse piezoelectric effect. With the development of wearable electronics, more and more electronics are increasingly integrated into clothing, either for functional or fashion reasons. This requires the next generation of energy harvesters moves into wearable electronics, which needs the magnetic and piezoelectric materials to be flexible and lightweight. Onuta *et al.*, have fabricated an electromagnetic energy harvester consisting of a magnetostrictive $Fe_{0.7}Ga_{0.3}$ thin film and a $Pb(Zr_{0.52}Ti_{0.48})O_3$ piezoelectric thin film on a 3.8-μm-thick Si cantilever, as shown in Fig. 4 [38]. The dependence of output voltage and harvested power on the AC magnetic field is shown in Fig. 20 [38]. The harvested peak power of 0.7 mW/cm$^3$ at 1 Oe is about 6 times larger than the value reported in the Terfenol-D/PZT/Terfenol-D laminate structures [76,77], which promotes the development of flexible energy harvesting materials.

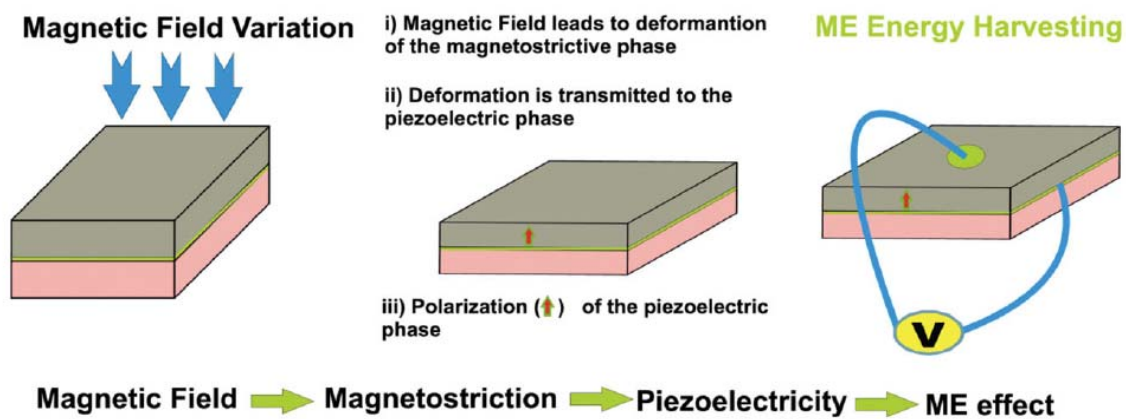

**Fig. 19.** Schematic show of harvesting mechanism for multiferroic materials. [39]

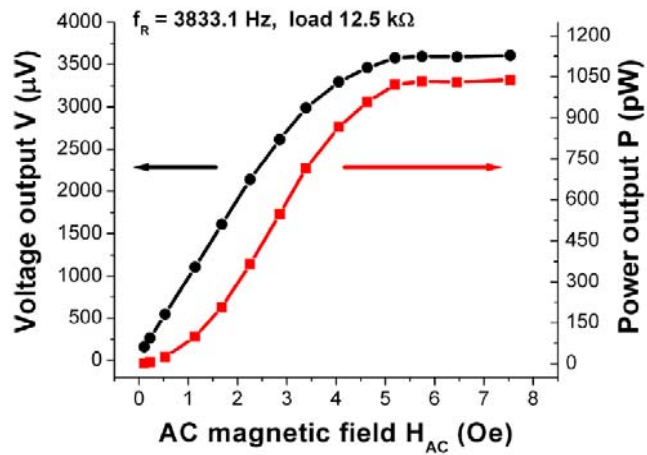

**Fig. 20.** Dependence of the output voltage and harvested power on the AC magnetic field of Pb($Zr_{0.52}Ti_{0.48}$)$O_3$/ $Fe_{0.7}Ga_{0.3}$ cantilevers [38]

**4.3 Flexible polymer based magnetic composite applied in actuators**

Polymer-based magnetic composite are flexible, lightweight, and easily processed, which are widely applied in MEMS [78]. The magnetically actuated micro-devices can be controlled without wire as long as the actuation environment is magnetically transparent, and therefore can be operated in air, vacuum, water, etc [79]. The flexible, wireless control of micro-devices makes polymer based magnetic composite attractive for many applications. However, the actuation or precise control of the polymeric components is a difficult issue. Lee *et al.*, have prepared an actuator consisting of a body plate of 50 μm thickness and four legs using photoresist SU-8 [80]. Three square sections of NiFe films of 10 μm thickness are in turn electroplated on the top and bottom sides of the SU-8 body, as shown in Fig. 21 [80]. When the magnetic field is applied to the actuator along with longitudinal direction, the magnetostriction of NiFe films make the actuator curve and move along the field direction, which is useful in the micro-machine area such as magnetic field driven drug delivery. Kim *et al.*, have presented a new magnetic polymeric micro-actuator, which allows the programming of heterogeneous magnetic anisotropy at the microscale [81]. As seen in Fig. 22, the micro-actuator is composed of four magnetic bodies having different magnetic easy axes, such that it has various configurations according to the applied external magnetic field direction. By freely programming the rotational axis of each component, the polymeric micro-actuators can undergo predesigned, complex two- and three-dimensional motions.

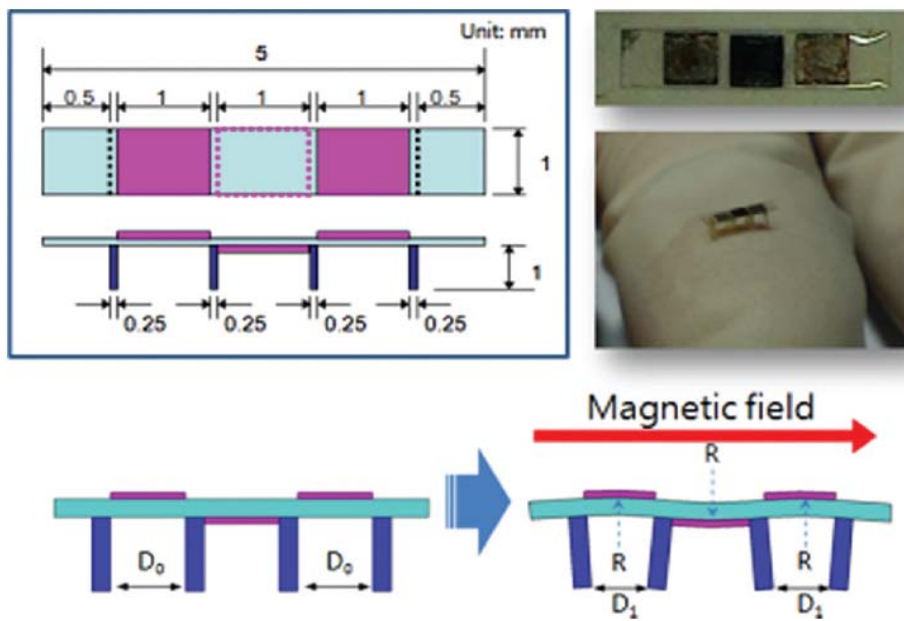

**Fig. 21.** Schematic view with photograph of a fabricated NiFe thin film worm actuator and its deflected motion. [80]

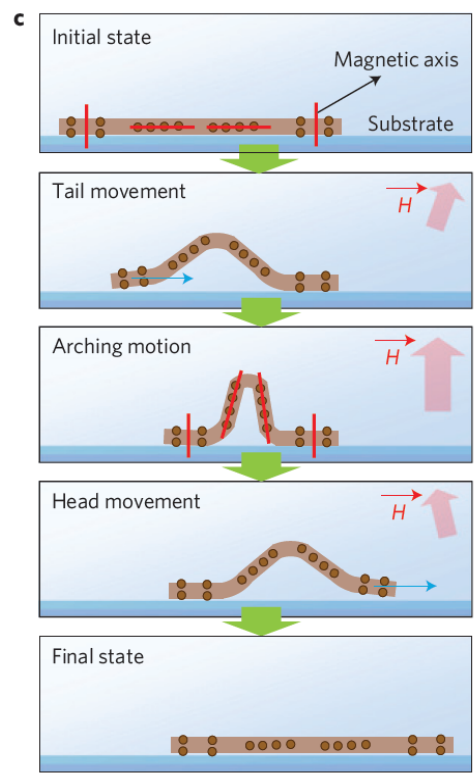

**Fig. 22.** Movement of polymeric micro-actuators by magnetic field. [81]

### 4.4 Flexible soft magnetic films applied in microwave devices

Flexible soft magnetic films exhibit high microwave permeability, which are of practical importance for a number of applications, such as high frequency inductors, transformers, shielding, and electromagnetic interference (EMI) devices [82,83,84]. Additionally, the flexible films can be cut easily and be used on arbitrary surface. Due to the magnetostriction effect, the magnetic anisotropy of flexible soft magnetic films can be tuned by changing the status of strain in flexible films, which provides the possibility to overcome the Snoek's limits and design the frequency tunable microwave devices [85,86]. Flexible thin films of magnetic alloys, such as FeCoB, FeCoBSi, FeTaN, FeZrN, and CoAlO, and the polymer-nanoparticle composites have been prepared due to their high frequency applications [87,88,89,90,91,92]. Zuo *et al.*, have prepared [Fe–Co–Si ($d$)/native oxide]$_{50}$ multilayer films with different metallic layer thicknesses ($d$) on flexible Kapton substrates by DC magnetron sputtering [93]. Figure 23 depicts the permeability spectra for the multilayer films with various $d$. The films exhibit relatively high values of complex permeability and ferromagnetic resonance frequency up to 7.9 GHz, indicating the great potential for applications in high-frequency electromagnetic devices. Rasoanoavy, *et al.*, have prepared a flexible CoFeB/PVDF/CoFeB composite material and observed a 30% variation in the microwave permeability under the application of a 1.5 MV/m electric field, due to the modified magnetic anisotropy caused by the strain-mediated magnetoelectric effect [94].

Employing the exchange bias effect is another way to promote the resonance frequency of the magnetic films and devices. Phuoc *et al.*, have prepared Permalloy-FeMn multilayers on flexible Kapton substrates [95]. A multiple-stage magnetization reversal and consequently plural ferromagnetic resonance absorption has been observed, which is possibly interpreted in terms of the different exchange interfacial energy acting on each layers. Based on these results, they have demonstrated a wide-band microwave absorber by using flexible Permalloy-FeMn multilayers. Figure 24 shows the frequency dependence of the reflection loss of flexible Permalloy-FeMn multilayers. The working bandwidth (the absorption width where the reflection loss is less than 10 dB) of the present film is rather broad ranging from 1 to 4 GHz, indicating that flexible exchange-biased multilayer systems are promising for future high-frequency applications.

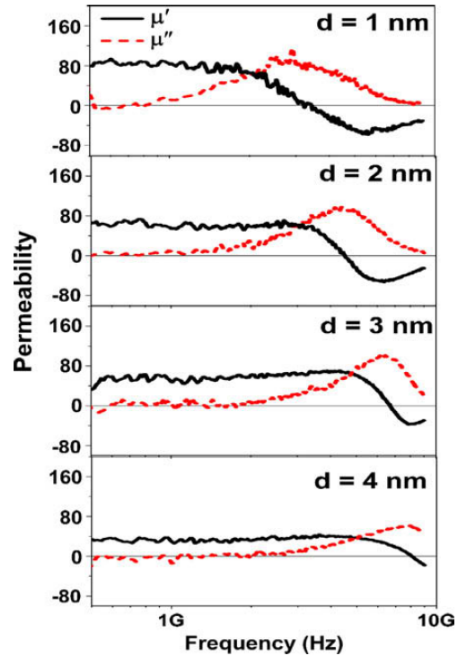

**Fig. 23.** Permeability spectra of [Fe–Co–Si (*d*)/native oxide]$_{50}$ multilayer films. [93]

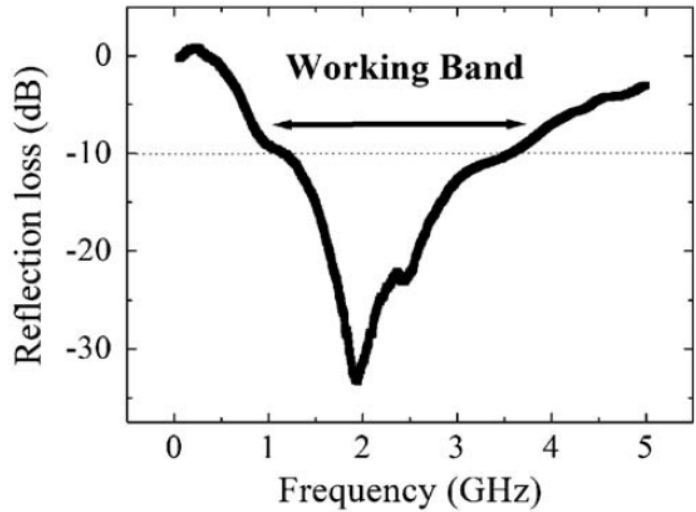

**Fig. 24.** Frequency dependence of the reflection loss of flexible Permalloy–FeMn multilayers.[95]

## 5. Conclusions and perspectives

    Investigations on flexible magnetic films and their applications are new-arising areas, which show potential importance for the development of flexible electronics. In order to give a comprehensive understanding of the flexible magnetic films and related applications, we have reviewed recent advances in the study of flexible magnetic films including the fabrication methods, the physical properties, and the applications of the flexible magnetic

films and devices. (a) By deposition of magnetic films on the flexible substrates, transfer and bonding approach, or removing sacrificial layers, one can prepare the desired flexible magnetic films. Generally, all flexible magnetic films require fabricated on a supporting flexible substrate. Therefore, flexible substrates with good flatness, high treatment temperature, stable thermal stability, good mechanical properties, etc, are very important for the fabrication of high quality flexible magnetic films. (b) Due to the flexibility of flexible magnetic films, the strain effect on the magnetic-related properties is very important both for the fundamental research and the practical applications. The magnetic anisotropy, exchanged bias, coercivity, and magnetoresistance of flexible magnetic films and devices can be effectively manipulated by applying external strains, which has a good potential application in the straintronic devices. However, in order to realize the multifunctional straintronic devices, the further comprehensive studies on the physical properties of flexible magnetic films are necessary, especially under multi-field conditions, like electric field, magnetic field, and strain field, etc. (c) Flexible magnetic films can be applied in the spintronic devices, energy harvester, actuators, and microwave devices, etc, due to the properties of bio-compatible, low-cost, light-weight, and compact.

**References:**

1. Gosele U and Tong Q Y 1998 *Annu. Rev. Mater. Sci.* **28** 215
2. Rogers J A, Someya T and Huang Y G 2010 *Science* **327** 1603
3. Kim D H and Rogers J A 2008 *Adv. Mater.* **20** 4887
4. Crabb R L and Treble F C 1967 *Nature* **213** 1223
5. Ray K A 1967 *IEEE Trans. Aerosp. Electron.* Syst. **AES-3(1)** 107
6. Yang N D, Harkin G, Bunn R M, McCulloch D J, Wilks RW and Knapp A G 1997 *IEEE Electron. Device Lett.* **18** 19
7. Smith P M, Carey P G and Sigmon T W 1997 *Appl. Phys. Lett.* **70** 342
8. Someya T, Sekitani T, Iba S, Kato Y, Kawaguchi H and Sakurai T 2004 *Proc. Natl. Acad. Sci. USA* **101** 9966
9. Forrest S R 2004 *Nature* **428** 911
10. Rogers J A, Bao Z, Baldwin K, Dodabalapur A, Crone B, Raju V R, Kuck V, Katz H, Amundson K, Ewing J and Drzaic P 2001 *Proc. Natl. Acad. Sci. USA* **98** 4835
11. Kanti J, Klosner M, Zemel M and Raghunandan S 2005 *Proceedings of the IEEE* **93** 1500
12. Dobrzański L A, Drak M and Ziębowicz B 2006 *J. Achievem. Mater. Manufact. Eng.* **17** 37
13. Osaka T, Takai M, Hayashi K, Ohashi K, Saito M A and Kazuhiko Yamada 1998 *Nature* **392**, 796
14. McHenry M E, Willard M A and Lauglin D E 1999 *Prog. Mater. Sci.* **44** 291
15. Fastenau H J and Loenen E J 1996 *J. Magn. Magn. Mater.* **1570158** 1
16. OlabiA AG and Grunwald A 2008 Mater. Design **29** 469
17. Thompson D A, R L T and Mayadas A 1975 *IEEE Trans. Magn.* **11** 1039
18. Wolf S A, Awschalom D D, Buhrman R A, Daughton J M, von Molnar S, Roukes M L, Chtchelkanova A Y and Treger D M 2001 *Science* **294** 1488
19. Fert A 2008 *Rev. Mod. Phys.* **80** 1517
20. Huang G S and Mei Y F 2012 *Adv. Mater.* **24** 2517
21. Melzer M, Makarov D, Calvimontes A, Karnaushenko D, Baunack S, Kaltofen R, Mei Y. F and Schmidt O G 2011 *Nano. Lett.* **11**, 2522
22. Choi M C, Kim Y K and Ha C S 2008 *Prog. Polym. Sci.* **33** 581
23. Eder F, Klauk H, Halik M, Zschieschang U, Schmid G, and Dehm C 2004 *Appl. Phys. Lett.*

**84** 2673

[24] Chang W Y, Fang T H and Lin Y C 2008 *Appl. Phys. A* **92** 693

[25] Bouziane K, Al Rawas A D, Maaz M and Mamor M 2006 *J. Alloys Compounds* **414** 42

[26] Breidbach M, Burgler D E and Grunberg P 2006 *J. Magn. Magn. Mater.* **307** L1

[27] Tian Z C, Sakaue K and Terauchi H 1994 *J. Appl. Phys.* **76** 3899

[28] Dai G H, Zhan Q F, Liu Y W, Yang H L, Zhang X S, Chen B and Li R W 2012 *Appl. Phys. Lett.* **100** 122407

[29] Zhang X S, Zhan Q F, Dai G H, Liu Y W, Zuo Z H, Yang H L, Chen B and Li R W 2013 *J. Appl. Phys.* **113** 17A901

[30] Parkin S S P 1996 *Appl. Phys. Lett.* **69** 3092

[31] Chen Y F, Mei Y F, Kaltofen R, Monch J I, Schuhmann J, Freudenberger J, Klauβ H J and Schmidt O G 2008 *Adv. Mater.* **20** 3224

[32] Oh S J, Jadhav M, Lim J, Reddy V, Kim C G 2013 *Biosens. Bioelectron.* **41** 758

[33] Melzer M, Lin G, Makarov D and Schmidt O G 2012 *Adv. Mater.* **24** 6468

[34] Zhao S F, Wan J G, Yao M L, Liu J M, Song F Q and Wang G H 2010 *Appl. Phys. Lett.* **97** 212902

[35] Liang J J, Xu Y F, Sui D, Zhang L, Huang Y, Ma Y F, Li F F and Chen Y S 2010 *J. Phys. Chem. C* **114** 17465

[36] Bell D J, Lu T J, Fleck N A and Spearing S M 2005 *J. Micromech. Microeng.* **15** S153

[37] Kim Y S, Jang S, Lee C S, Jin W Hg, Cho Il J, Ha M H, Nam H J, Bu J U, Chang S Il and Yoon E 2007 *Sens. Actuators A* **135** 67

[38] Onuta T D, Wang Y, Long C J and Takeuchi 2011 *Appl. Phys. Lett.* **99** 203506

[39] Martins P and Lanceros-Méndez S 2013 *Adv. Funct. Mater.* DOI:10.1002/adfm.201202780

[40] Inoue S, Utsunomiya S, Saeki T and Shimoda T 2002 *IEEE Trans. Electron. Devices* **49** 1353

[41] Lee Y, Li H and Fonash S J 2003 *IEEE Electron. Device Lett.* **24** 19

[42] Reese C, Roberts M, Ling M M and Bao Z N 2004 *Materials Today* **ISSN:1369 7021** (Elsevier Ltd.)

[43] Meitl M A, Zhu Z T, Kumar V, Lee K J, Feng X, Huang Y Y, Adesida I, Nuzzo R G and Rogers J A 2006 *Nat. Mater.* **5** 33

[44] Lee Y, Bae S, Jang H, Jang S, Zhu S, Sim S H, Song Y I, Hong B H and Ahn J H 2010

   *Nano Lett.* **10** 490

[45] Donolato M, Tollan C, Porro J M, Berger A and Vavassori P 2013 *Adv. Mater.* **25** 623

[46] Linder V, Gates B D, Ryan D, Parviz B A and Whitesides G M 2005 *small* **1** 730

[47] Buhlery J, Steiner F P and Baltes H 1997 *J. Micromech. Microeng.* **7** R1

[48] Heczko O, Thomas M, Niemann R, Schultz L and Fähler S 2009 *Appl. Phys. Lett.* **94** 152513

[49] Tillier J, Bourgault D, Barbara B, Pairis S, Porcar L, Chometon P, Dufeu D, Caillault N and Carbone 2010 *J Alloys Compd.* **489** 509

[50] Bechtold C, Buschbeck J, Lotnyk A, Erkartal B, Hamann S, Zamponi C, Schultz L, Ludwig A, Kienle L, Fahler S and Quandt E 2010 *Adv. Mater.* **22** 2668

[51] Edler T and Mayr S G 2010 *Adv. Mater.* **22** 4969

[52] Backen A, Yeduru S R, Kohl M, Baunack S, Diestel A, Holzapfel B, Schultz L, Fahler S 2010 *Acta Mater.* **58** 3415

[53] Zuo Z H, Chen B, Zhan Q F, Liu Y W, Yang H L, Li Z X, Xu G J and Li R W 2012 *J. Phys. D: Appl. Phys.* **45** 185302

[54] Shin J, Kim S H, Suwa Y, Hashi S and Ishiyama K 2012 *J. Appl. Phys.* **111**, 07E511

[55] Cullity B D 1972 *Introduction to Magnetic Materials (Addison-Wesley, Reading)*

[56] Zhang X S, Zhan Q F, Dai G H, Liu Y W, Zuo Z H, Yang H L, Chen B and Li R W 2012 *Appl. Phys. Lett.* **100** 022412

[57] Chen Y F, McCord J, Freudenberger J, Kaltofen R and Schmidt O G 2009 *J Appl. Phys.* **105**, 07C302

[58] Parkin S S P, Roche K P and Suzuki T 1992 *Jpn. J. Appl. Phys.* **31** L1246

[59] Barraud C, Deranlot C, Seneor P, Mattana R, Dlubak B, Fusil S, Bouzehouane K, Deneuve D, Petroff F and Fert A 2010 *Appl. Phys. Lett.* **96** 072502

[60] Baibich M N, Broto J M, Fert A, Nguyen F, Dau V, Petroff F, Eitenne P, Creuzet G, Friederich A and Chazelas J 1988 *Phys. Rev. Lett.* **61** 2472

[61] Florez S H and Gomez R D 2003 *IEEE Trans. Magn.* **39** 3411

[62] Dokupil S, Bootsmann M T, Stein S, Lohndorf M and Quandt E 2005 *J. Magn. Magn. Mater.* **291** 795

[63] Özkaya B, Saranu S R, Mohanan S and Herr U 2008 *phys. stat. sol. (a)* **205** 1876

[64] Salata O V 2004 *J. Nanobiotechnol.* **2** 3

⁶⁵ Yu J, Huang D Y, Muhammad Z Y, Hou Y and Gao S 2013 *Chin. Phys. B* **22** 027506

⁶⁶ Nam J M, Thaxton C C, Mirkin C A 2003 *Science* **301** 1884

⁶⁷ Parkin S S P 1995 *Annu. Rev. Mater. Sci* **25** 257

⁶⁸ Pannetier M, Fermon C, Goff G L, Simola J and Kerr E 2004 *Science* **304** 1648

⁶⁹ Monch I, Makarov D, Koseva R, Baraban L, Karnaushenko D, Kaiser C, Arndt K F and Schmidt O G 2011 *ACS Nano* **5** 7436

⁷⁰ Melzer M, Karnaushenko D, Makarov D, Baraban L, Calvimontes A, Monch I, Kaltofen R, Mei Y F and Schmidt O G 2012 *RSC Advances* **2** 2284

⁷¹ Ma Z Q 2011 *Science* **333** 830

⁷² Karnaushenko D, Makarov D, Yan C L, Streubel R and Schmidt O G 2012 *Adv. Mater.* **24** 4518

⁷³ Chalasani S and Conrad J M 2008 *Southeastcon IEEE* 442-447

⁷⁴ Stanton S C, McGehee, C C and Mann Brian P 2009 *Appl. Phys. Lett.* **95** 174103

⁷⁵ Qi Y and McAlpine M C 2010 *Energy Environ. Sci.* **3** 1275

⁷⁶ Zhang C L, Yang J S and Chen W Q 2009 *Appl. Phys. Lett.* **95** 1

⁷⁷ Li P, Wen Y, Liu P, Li X, Jia C 2010 *Sens. Actuators A* **157** 100

⁷⁸ Winey K I and Vaia R A 2007 *MRS Bull.* **32** 314

⁷⁹ Cugat O, Delamare J and Reyne G 2003 *IEEE Trans. Magn.* **39** 3607

⁸⁰ Lee H S, Cho C D, Zehnder A T and Choi K W 2011 *J Appl. Phys.* **109**, 07E501

⁸¹ Kim J Y, Chung S E, Choi S E, Lee H W, Kim J H and Kwon S H 2011 *Nat. Mater.* **10** 747

⁸² Ha N D, Phan M H and Kim C O 2007 *Nanotechnology* **18** 155705

⁸³ Qiao L, Wen F, Wei J, Wang J and Li F 2008 *J. Appl. Phys.* **103** 063903

⁸⁴ Phuoc N N, Xu F and Ong C K 2009 *Appl. Phys. Lett.* **94** 092505

⁸⁵ Acher O and Dubourg S 2008 *Phys. Rev.* B **77** 104440

⁸⁶ Iakubov I T, Lagarkov A N, Maklakov S A, Osipov A V, Rozanov K N, Ryzhikov I A and Starostenko S N 2006 *J. Magn. Magn. Mater.* **300** 74

⁸⁷ Phuoc N N, Hung L T and Ong C K 2010 *J. Phys. D: Appl. Phys.* **43** 255001

⁸⁸ Lu H P, Yang J and Deng L J 2010 *Piers Online* **6** 105

⁸⁹ Liu Z W, Liu Y, Yan L, Tan C Y and Ong C K 2006 *J. Appl. Phys.* **99** 043903

⁹⁰ Liu Z W, Zeng D C, Ramanujan R V and Ong C K 2010 *J. Appl. Phys.* **107** 09A505

[91] Ma Y G and Ong C K 2007 *J. Phys. D: Appl. Phys.* **40** 3286

[92] Stojak K, Pal S, Srikanth H, Morales C, Dewdney J, Weller T and Wang J 2011 *Nanotechnology* **22** 135602

[93] Zuo H P, Ge S H, Wang Z K, Xiao Y H and Yao D S 2010 *Scripta Mater.* **62** 766

[94] Rasoanoavy F, Laur V, Blasi S D, Lezaca J, Quéffélec P, Garello K and Viala B 2010 *J. Appl. Phys.* **107** 09E313

[95] Phuoc N N, Xu F, Ma Y G, Ong C K 2009 *J. Magn. Magn. Mater.* **321** 2685